\documentclass[a4paper,twoside]{article}
\usepackage{amssymb}
\usepackage{amsmath}
\usepackage{amsfonts}
\usepackage{graphicx}

\begin{document}

\title{\Large{\bf 
Black hole interior from loop quantum gravity
}}
\author{\\ Leonardo Modesto
 \\[1mm]
\small{ Department of Physics, Bologna University V. Irnerio 46, I-40126 Bologna \& INFN Bologna, EU}
   }

\date{\ } 
\maketitle

\begin{abstract}
In this paper we calculate modifications to the Schwarzschild 
solution by using a semiclassical analysis of loop quantum black hole.
We obtain a metric inside the event horizon that coincides with
the Schwarzschild solution near the horizon but that is substantially 
different at the Planck scale. In particular we obtain 
a bounce of the $S^2$ sphere for a minimum value of the radius
and that it is possible to have another event
horizon close to the $r=0$ point.

\end{abstract}

\section*{Introduction}
Quantum gravity, the theory that wants reconcile general relativity and quantum
mechanics, is one of  major problem in theoretical physics today.
General relativity tells as that because also the space-time is dynamical, 
it is not possible to study other interactions on a fixed background.
The background itself is a dynamical field.
 
 Among the quantum gravity theories, the theory called
 ``loop quantum gravity" \cite{book}
is the most widespread nowadays.
This is one of the non perturbative and 
background independent approaches to quantum gravity
(another non perturbative approach to quantum gravity is called  ``asymptotic safety 
quantum gravity" \cite{MR}). 
In the last years 
the applications of loop quantum gravity ideas to minisuperspace models 
lead to some interesting results to solve  
the problem of space-like singularity in quantum gravity. 
As shown in cosmology \cite{Boj}, \cite{MAT} and recently in black hole physics
\cite{work1}, \cite{work2}, \cite{work3}, \cite{ABM} it is 
possible to solve the cosmological singularity problem and the
black hole singularity problem by using 
the tools and ideas developed in full loop quantum gravity theory. 
In the other well known approach to quantum gravity,
the called  ``asymptotic safety quantum gravity",
 authors \cite{BR}, using 
the $G_N$ running coupling constant 
obtained in  ``asymptotic safety quantum gravity",
have showed that non perturbative 
quantum gravity effects give a much less singular Schwarzschild metric
and that for particular values of the black hole mass 
it is possible to have the formation of another 
event horizon.

In this paper we study the space-time inside the event horizon
at the semiclassical level using a constant polymeric parameter $\delta$ 
(see the paper \cite{NC} for an analysis of the black hole interior using
a non constant polymeric parameter).
We consider the Hamiltonian constraint obtained in \cite{ABM};
in particular we study the Hamiltonian constraint introduced in
the first paper of reference \cite{ABM}, where the authors have taken 
the general version of the constraint for real values of the 
Immirzi parameter $\gamma$. 

This paper is organized as follows.  In the first section we briefly
recall the Schwarzschild solution inside the event horizon ($r<2M G_N$)
of \cite{ABM}. In the second section we introduce the Hamiltonian 
constraint in terms of holonomies and then the relative trigonometric
form solving the Hamilton equations of motion.
In the third section we give the metric form of the solution 
and we discuss the new physics suggested by loop quantum gravity.

\section{Schwarzschild solution inside the event horizon\\ in Ashtekar variables}
We recall the classical Schwarzschild solution inside the event 
horizon \cite{ABM}.
For the homogeneous but non isotropic Kantowski-Sachs
space-time the Ashtekar's connection and density triad 
are (after the fixing of a residual global $SU(2)$ gauge symmetry on the spherically reduced 
phase space \cite{ABM}) 
\begin{eqnarray}
&& A= c \tau_3 d x + b \tau_2 d \theta - b \tau_1 \sin \theta d \phi + \tau_3 \cos \theta d \phi,
\nonumber \\
&&E = p_c \tau_3 \sin \theta \frac{\partial}{\partial x} + p_b \tau_2 \sin \theta \frac{\partial}{\partial \theta} - p_b \tau_1 \frac{\partial}{\partial \phi}.
\end{eqnarray}
The components variables in the phase space 
can be read 
From the symmetric reduced connection and density 
triad we can read the components variables in the phase space:
$(b, p_b)$, $(c, p_c)$.
The Poisson algebra is:
$\{c, p_c \} = 2 \gamma G_N$, $\{b, p_b \} = \gamma G_N$.
Following papers \cite{ABM} we recall that the classical
Hamiltonian constraint in terms of the components variables is  
\begin{eqnarray}
\mathcal{C}_{H} = - \frac{1}{2 \gamma G_N} \left[ (b^2 + \gamma^2) \frac{p_b}{b} + 2 c \, p_c \right],
\label{Ham1}
\end{eqnarray}
in the gauge $N=\gamma \, \mbox{sgn}(p_c) \sqrt{|p_c|}/16 \pi G_N b$. 
Hamilton equations of motion are 
\begin{eqnarray}
&& \dot{b} = \{ b, \mathcal{C}_{H} \} = - \frac{b^2 + \gamma^2}{2 b},
\hspace{2cm}
\dot{p_b} = \{ p_b, \mathcal{C}_{H} \} =  \frac{1}{2} \Big[p_b -\frac{\gamma^2 p_b}{b^2}\Big],
\nonumber \\
&&  \dot{c} = \{ c, \mathcal{C}_{H} \} = - 2 c, 
\hspace{2.9cm}
\dot{p_c} = \{ p_c, \mathcal{C}_{H} \} = 2 p_c.
\label{Eq.1}
\end{eqnarray}
Solutions of equations (\ref{Eq.1}) using the time parameter $t \equiv e^T$ 
and redefining the integration constant $\equiv e^{T_0} = m$ 
(see the first of papers in \cite{ABM}) are 
\begin{eqnarray}
&& b(t) = \pm \gamma \sqrt{2m/t - 1}, \hspace{2cm} p_b(t) = p_b^{(0)} \sqrt{t(2 m - t)} \nonumber \\
 && c(t) = \mp \gamma m t^{-2}, \hspace{3cm} p_c(t) = \pm t^2.
 \label{Sol.1}
\end{eqnarray}
This is exactly the Schwarzschild solution inside the event horizon as you
can verify passing to the metric form defined by $h_{ab} = \mbox{diag}(p_b^2/p_c,p_c, p_c \sin^2 \theta)$ ($m$ contains the gravitational constant parameter $G_N$).

\section{
Semiclassical dynamics from loop quantum gravity
}

We recall now the Hamiltonian constraint coming from  ``loop quantum
black hole `` \cite{ABM} in terms of the explicit trigonometric form of holonomies.
The Hamiltonian constraint depends explicitly on the parameter $\delta$ that 
defines the length of the curves along witch we integrate the connections to define
the holonomies \cite{ABM}. We use the notation $\mathcal{C}^{\delta}$ for the 
hamiltonian constraint to stress the dependence on the parameter $\delta$.
The Hamiltonian constraint in terms of holonomies is 
\begin{eqnarray}
&& \hspace{0.0cm} \mathcal{C}^{\delta} = 
 \frac{ - N}{(8 \pi G_N)^2 \gamma^3 \delta^3} \, 
   {\rm Tr} \Big[\sum_{ijk} \epsilon^{ijk} h_i^{(\delta)} h_j^{(\delta)} h_i^{(\delta) -1}
h_j^{(\delta) -1}
 h_k^{(\delta)}
\left\{h_k^{(\delta) -1}, V \right\}
 + 2 \gamma^2 \delta^2 \tau_3 h_1^{(\delta)}\left\{h_1^{(\delta) -1}, V\right\}     
    \hspace{-0.0cm} \Big]  \nonumber \\
&&\hspace{0.5cm} = - \frac{ N}{2 G_N \gamma^2 } 
\Bigg\{ 2 \frac{\sin \delta c}{\delta} \ \frac{\sin \delta b}{\delta} \ \sqrt{|p_c|} 
+ \left(\frac{\sin^2 \delta  b}{\delta^2} + \gamma^2  \right) \frac{p_b \ \mbox{sgn}(p_c)}{\sqrt{|p_c|}} \Bigg\}, 
\label{CH}
\end{eqnarray} 
where $V = 4 \pi \sqrt{|p_c|} p_b$ is the spatial section volume and we have
calculated the Poisson brackets using the symplectic structure given in the previous 
section. The holonomies are 
\begin{eqnarray}
&&  h_1^{\delta} = \cos \frac{\delta c}{2} + 2 \tau_3 \sin \frac{\delta c}{2}, \nonumber \\
&&  h_2^{\delta} = \cos \frac{\delta b}{2} - 2 \tau_1 \sin \frac{\delta b}{2}, \nonumber \\
&& h_3^{\delta} = \cos \frac{\delta b}{2} + 2 \tau_2 \sin \frac{\delta b}{2}. 
\end{eqnarray}
Now we can solve exactly the new Hamilton equations of motion if we take a gauge
where the equations for the canonical pairs $(b, p_b)$ and $(c, p_c)$ are decoupled.
A useful gauge is 
$N = \frac{\gamma \sqrt{|p_c|} \mbox{sgn}(p_c) \delta^2}{ \sin \delta b}$ and in this 
particular gauge the Hamiltonian constraint becomes 
\begin{eqnarray}
\mathcal{C}^{\delta} = - \frac{1}{2 \gamma G_N} 
\Big\{ 2 \sin \delta c  \  p_c +
\Big(\sin \delta b + \frac{\gamma^2 \delta^2}{\sin \delta b} \Big) 
p_b \Big\}.
\label{FixN}
\end{eqnarray}
From (\ref{FixN}) we obtain two independent set of equations of motion on the 
phase space 
\begin{eqnarray}
&& \dot{c} = - 2 \sin \delta c, \hspace{3cm} \dot{p_c} = 2 \delta p_c \cos \delta c 
\nonumber \\
&&\dot{b} = - \frac{1}{2} \Big(\sin \delta b + \frac{ \gamma^2 \delta^2}{\sin \delta b} \Big), \hspace{1.2cm}
\dot{p_b} =  \frac{\delta}{2} \, \cos \delta b \Big(1 - \frac{ \gamma^2 \delta^2}{\sin^2 \delta b} \Big) p_b.
\end{eqnarray}  
Solving the first two equations
for $c(T)$ and $p_c(T)$ we obtain 
\begin{eqnarray}
&& c(T) =  \frac{2}{\delta} \arctan \Big( \mp \frac{\gamma \delta m p_b^{(0)}}{2} e^{- 2 \delta T} \Big),
\nonumber \\
&& p_c (T) = \pm e^{- 2 \delta T} \Big[\Big(\frac{\gamma \delta m p_b^{(0)}}{2}\Big)^2  + e^{4 \delta T} \Big].
\label{Sol.cpc}
\end{eqnarray}
Introducing a new time parametrization $t \equiv e^{\delta T}$ we obtain 
\begin{eqnarray}
 && c(t) =  \frac{2}{\delta} \arctan \Big( \mp \frac{\gamma \delta m p_b^{(0)}}{2 t^2}  \Big) \,\,
\rightarrow \,\,  \mp \frac{\gamma m p_b^{(0)}}{ t^2} 
\nonumber \\
&& p_c (t) = \pm 
\frac{1}{t^2} 
 \Big[\Big(\frac{\gamma \delta m p_b^{(0)}}{2}\Big)^2  + t^4 \Big] \,\, \rightarrow \pm t^2.
\label{Sol.cpc2}
\end{eqnarray}
In (\ref{Sol.cpc2}) we have calculated the small $\delta$ limit for 
the solution $c(t)$ and $p_c(t)$, obtaining the 
Schwarzschild solution of paragraph one in equation (\ref{Sol.1}) and calculated in \cite{ABM}.
A substantial difference between the Schwarzschild solution and 
the solution (\ref{Sol.cpc2}) is that in the second case there is an
absolute minimum in $t_{min} = (\gamma \delta m p_b^{(0)}/2)^{1/2}$,
where $p_c$ assume the value $p_{c}(t_{min}) = \gamma \delta m p_b^{(0)}>0$.
In the next section we will analyze the new physics coming from loop quantum gravity 
Hamiltonian constraint. 

At this point we integrate the equation of motion for $b(t)$ 
obtaining the following solution (we write the solution in the time coordinate $t$)
\begin{eqnarray}
\cos \delta b = \sqrt{1 + \gamma^2 \delta^2} 
\left[ \frac{\sqrt{1 + \gamma^2 \delta^2} + 1 - \Big(\frac{ 2 m}{t} \Big)^{\sqrt{1 + \gamma^2 \delta^2}}
(\sqrt{1 + \gamma^2 \delta^2} - 1)}
{\sqrt{1 + \gamma^2 \delta^2} + 1 + \Big(\frac{ 2 m}{t} \Big)^{\sqrt{1 + \gamma^2 \delta^2}}
(\sqrt{1 + \gamma^2 \delta^2} - 1)}
\right].
\label{Sol.b}
\end{eqnarray}
To calculate $p_b(t)$ we introduce the solutions $c(t), p_c(t), b(t)$ in the Hamiltonian constraint
and we obtain $p_b(t)$ from the algebraic constraint  
equation $C^{\delta} = 0$. The solution of this equation gives $p_b(t)$ as
function of the other phase space functions, 
\begin{eqnarray}
p_b(t) = -  \frac{2 \ \sin \delta c \
\sin \delta b \ p_c }{\sin^2 \delta b + \gamma^2 \delta^2}.
\label{Sol.pb}
\end{eqnarray}
To obtain the explicit form of $p_b(t)$ in terms of the time coordinate $t$ 
it is sufficient to introduce in (\ref{Sol.pb}) the solution $\cos \delta b$ calculated in (\ref{Sol.b}).

We note that the solution is homogeneous until it is 
satisfied the trigonometric property $\cos \delta b\geqslant - 1$.
Using (\ref{Sol.b}) we can calculate the 
variable $t$ value (we define this $t^{\ast}$) until the 
solution is of Kantowski-Sachs type and we obtain 
\begin{eqnarray}
t^{\ast} = 2 m \left(\frac{\sqrt{1 + \gamma^2 \delta^2} - 1}{\sqrt{1 + \gamma^2 \delta^2} + 1}\right)^{\frac{2}{\sqrt{1 + \gamma^2 \delta^2}}}.
\end{eqnarray}
However we are interested in the semiclassical limit of the solution defined 
by $\delta \ll 1$, then in this particular limit $t^{\ast} \sim 0$ (see also the
next section). 

\begin{figure}
 \begin{center}
  \includegraphics[height=6cm]{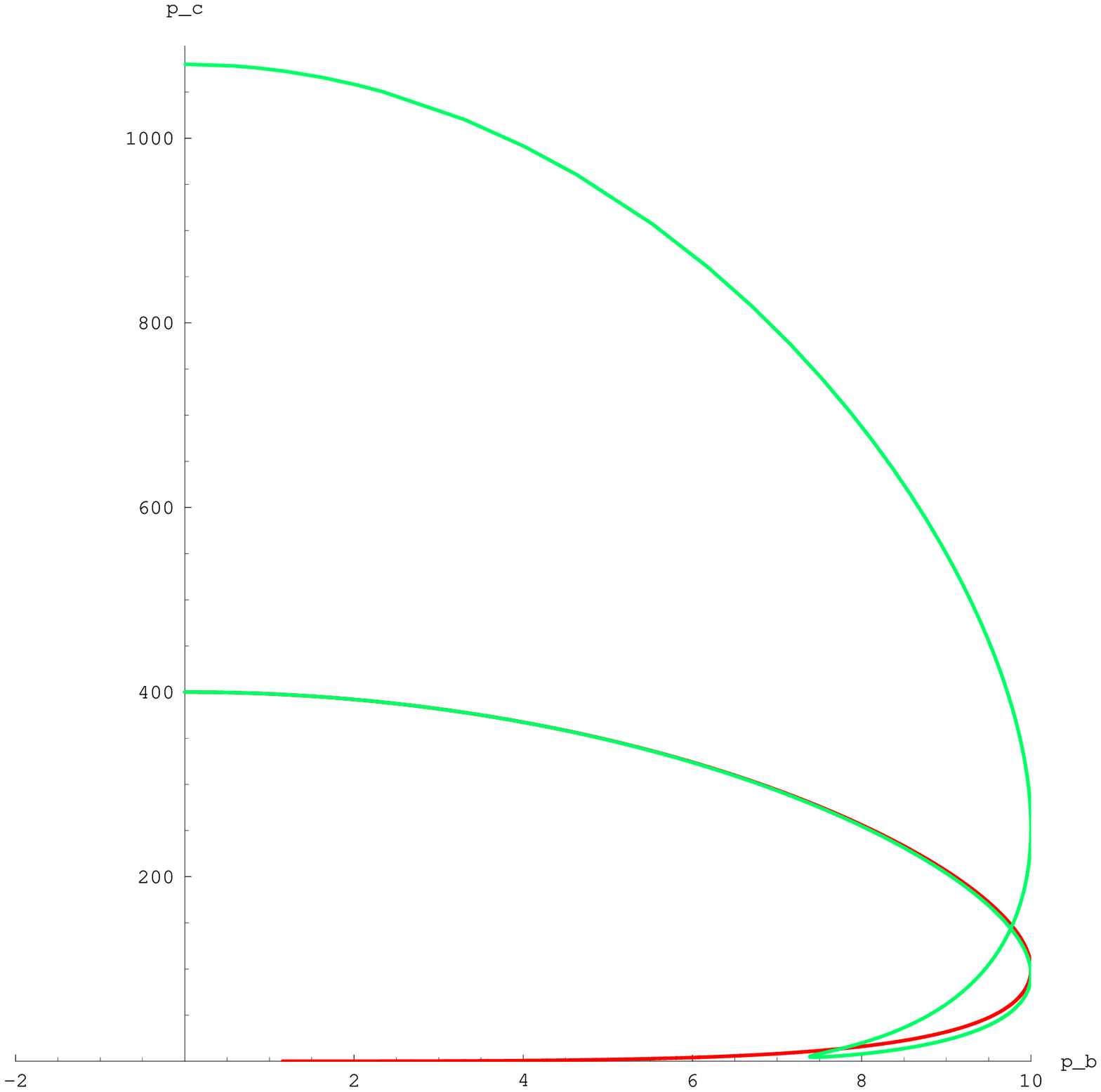}
   \includegraphics[height=6cm]{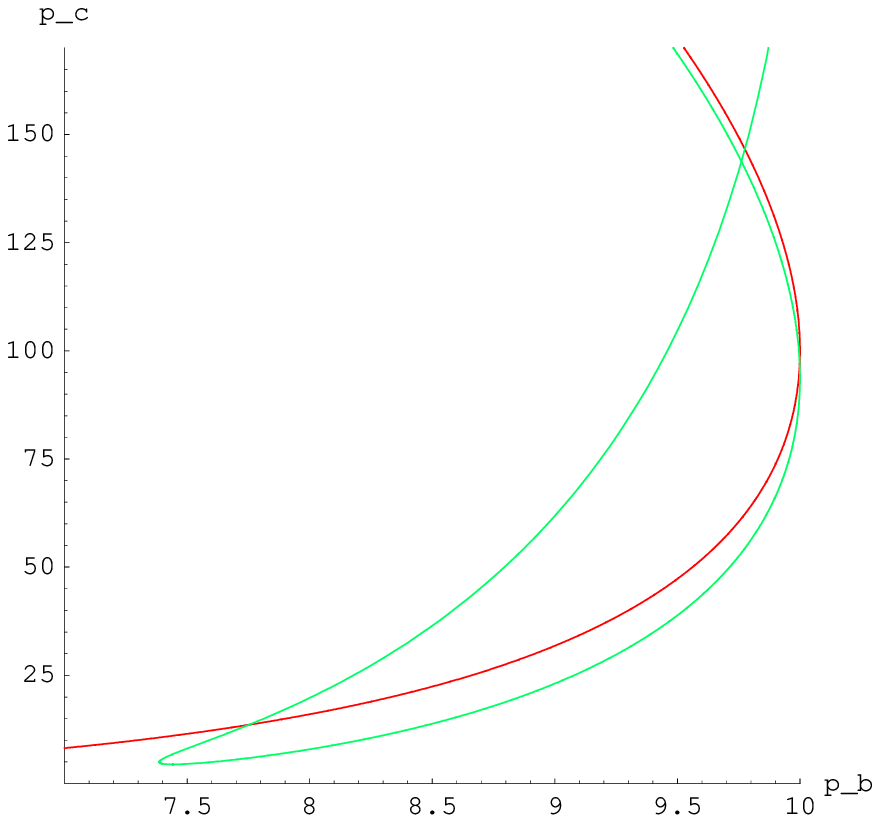}
  \end{center}
  \caption{\label{pbpc} 
   Semiclassical dynamical trajectory in the plane $p_b - p_c$
   The plots for $p_c > 0$ and for $p_c < 0$ are disconnected and symmetric
   but we plot only the positive values of $p_c$.  
  The red trajectory correspond to
                 the classical Schwarzschild solution and the 
                 green trajectory correspond to the semiclassical solution 
                (the green and red curves are continuum curves).
                In the plot on the right we have enlarged the region near the $p_b$ axis.}
  \end{figure}

Following \cite{ABM} we study the trajectory on 
the plane $p_c-p_b$ and we compare the result with
the Schwarzschild solution of the section one.
In Fig.\ref{pbpc}  we have a parametric plot of $p_c$ and $p_b$ (for $m = 10$) and $\gamma \delta \sim 1$ to amplify the quantum gravity effects in the plot (see the footnote 
in next section). We can observe the substantial difference with the classical case.
In the classical case (red line in Fig.\ref{pbpc})
 $p_c \rightarrow 0$ for $t\rightarrow 0$ and this point 
corresponds to the classical singularity. In the semiclassical case 
instead we start from $t=2m$ where $p_c \rightarrow (2m)^2$ and $p_b \rightarrow 0$ (this point 
corresponds to the Schwarzschild horizon) and decreasing $t$ we arrive to
a minimum value for $p_{c,m} \equiv p_c(t_{min}) >0$. From this point $p_c$
starts to grow another time until it assumes a maximum value for $p_b=0$
that corresponds to a new horizon in $t = t^{\ast}$ localized (see next section
where we study the metric form of the solution).
Our analysis refers to 
the region $t^{\ast} \leqslant  t  \leqslant 2 m$
and the plot in Fig.\ref{pbpc} refers to this time interval.  
The solution calculated is regular in the region $t^{\ast} \leqslant t \leqslant 2m$
in fact the co-triad $\omega$ \cite{ABM}, defined by (it is the inverse of the triad
$E$)
\begin{eqnarray}
 \omega = \frac{\mbox{sgn} (p_c) |p_b| \ \tau_3}{\sqrt{|p_c|}} d x +\mbox{sgn} (p_b) \ \sqrt{|p_c|} \ \tau_2 d \theta - \mbox{sgn} (p_b) \ \sqrt{|p_c|} \ \tau_1 \sin \theta d \phi, 
\label{co-triad}
\end{eqnarray}
is regular $\forall p_c$ in the region $t^{\ast} \leqslant t \leqslant 2m$.

\section{Metric form of the solution}
In this section we present the metric form of the solution and we give
a plot for any component of the 
Kantowski-Sachs metric 
$ds^2 = - N^2(t) dt^2 + X^2(t) dr^2 + Y^2(t)(d \theta^2 + \sin \theta d \phi^2)$.
We start recalling the relation between connection and metric variables
\begin{eqnarray}
Y^2(t) =|p_c(t)|, \,\,\,\,\,\,\, X^2(t) = \frac{p_b^2(t)}{|p_c(t)|}, \,\,\,\,\,\,\, 
N^2(t) = \frac{\gamma^2 \delta^2 |p_c(t)|}{(16 \pi G_N)^2 t^2 \sin^2 \delta b}. 
\label{metric}
\end{eqnarray}
\begin{figure}
 \begin{center}
  \includegraphics[height=6cm]{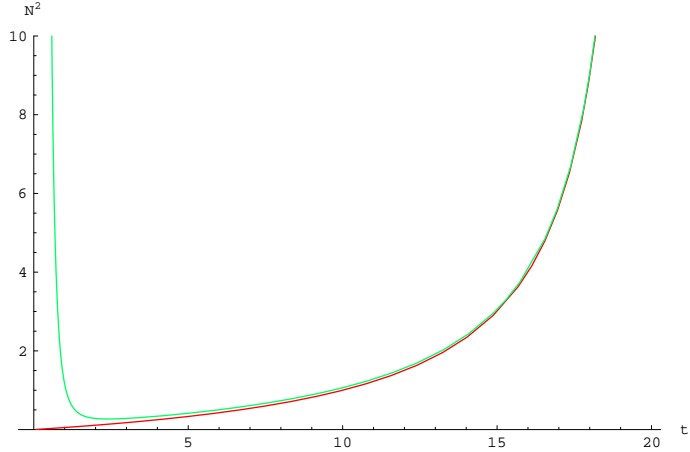}
  \end{center}
  \caption{\label{N2} Plot of the lapse function $N^2(t)$ 
 for $m = 10$ and $\gamma \delta \sim 1$
 (in the horizontal axis 
  we have the temporal coordinate $t$ and in the vertical axis the lapse function). 
                 The red trajectory correspond to
                 the classical Schwarzschild solution inside the event horizon and the 
                 green trajectory correspond to the semiclassical solution.}
  \end{figure}
We give now the explicit form of the metric components in terms of the temporal coordinate $t$.
The lapse function $N(t)$ is 
\begin{eqnarray}
(16 \pi G_N)^2 \ N^2(t) = \frac{\gamma^2 \delta^2 \left[ \left(\frac{\gamma \delta m}{2 t^2}\right)^2 +1 \right]}{
  1 - (1 + \gamma^2 \delta^2)
\left[ \frac{\sqrt{1 + \gamma^2 \delta^2} + 1 - \left(\frac{2m}{t} \right)^{\sqrt{1 + \gamma^2 \delta^2}}
(\sqrt{1 + \gamma^2 \delta^2} - 1)}
{\sqrt{1 + \gamma^2 \delta^2} + 1 + \left(\frac{2m}{t} \right)^{\sqrt{1 + \gamma^2 \delta^2}}
(\sqrt{1 + \gamma^2 \delta^2} - 1)}
\right]^2   
}.
\label{N2metric}
\end{eqnarray}
In Fig.\ref{N2} we have compared the classical Schwarzschild solution
inside the event horizon with the solution (\ref{N2metric})
for $m = 10$ and $\gamma \delta \sim 1$
(we have taken $\gamma \delta \sim 1$ to amplify, in the plot, the loop quantum gravity 
modifications at the Planck scale).
We can observe that the two
solutions are identically when we approach to the event horizon (which is in $t=20$ in the
units used in the plot) but are very different when we go toward $t\sim0$.
As we have explained in the previous section 
we consider the region 
$t > t^{\ast}$ and for $t = t^{\ast}$ the lapse function diverges, $N^2(t^{\ast}) \rightarrow +\infty$.
The semiclassical solution has a minimum before diverging in $t=t^{\ast}$.
In the classical solution instead (it is represented in red in Fig.\ref{N2}) $N^2(t)$ 
is very small for $t = t^{\ast}$ and goes to zero for $t \rightarrow 0$. 

The anisotropy function $X^2(t)$ is related to $p_b(t)$ and $p_c(t)$ by (\ref{metric}),
then introducing (\ref{Sol.pb}) and (\ref{Sol.cpc2}) in the second relation of (\ref{metric})
we obtain 
\begin{eqnarray}
X^2(t) = \frac{(2 \gamma \delta m)^2 \  
                         \left(1 - (1 + \gamma^2 \delta^2)
\left[ \frac{\sqrt{1 + \gamma^2 \delta^2} + 1 - \left(\frac{2m}{t} \right)^{\sqrt{1 + \gamma^2 \delta^2}}
(\sqrt{1 + \gamma^2 \delta^2} - 1)}
{\sqrt{1 + \gamma^2 \delta^2} + 1 + \left(\frac{2m}{t} \right)^{\sqrt{1 + \gamma^2 \delta^2}}
(\sqrt{1 + \gamma^2 \delta^2} - 1) }
\right]^2   \right) \  t^2 }{
(1 + \gamma^2 \delta^2)^2 \left(1 -
\left[ \frac{\sqrt{1 + \gamma^2 \delta^2} + 1 - \left(\frac{2m}{t} \right)^{\sqrt{1 + \gamma^2 \delta^2}}
(\sqrt{1 + \gamma^2 \delta^2} - 1)}
{\sqrt{1 + \gamma^2 \delta^2} + 1 + \left(\frac{2m}{t} \right)^{\sqrt{1 + \gamma^2 \delta^2}}
(\sqrt{1 + \gamma^2 \delta^2} - 1)}
\right]^2\right)^2
\Big[\Big(\frac{\gamma \delta m p_b^{(0)}}{2}\Big)^2  + t^4 \Big]
}.
\label{X}
\end{eqnarray}
\begin{figure}
 \begin{center}
  \includegraphics[height=6cm]{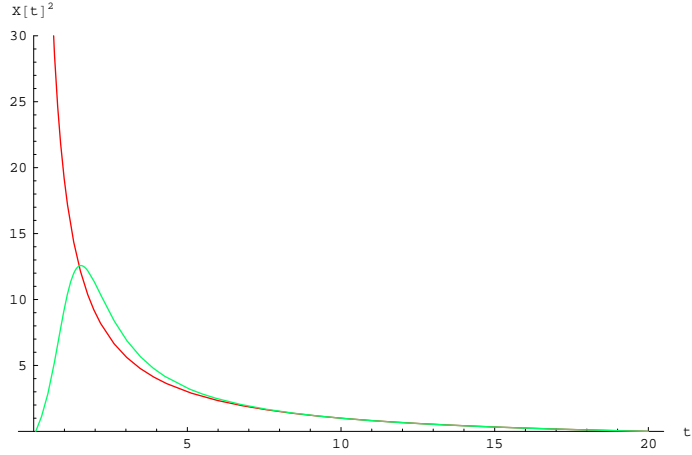}
  \end{center}
  \caption{\label{a2} 
   Plot of $X^2(t)$ 
 for $m = 10$ and $\gamma \delta \sim 1$
 (in the horizontal axis 
  we have the temporal coordinate $t$ and in the vertical axis we have $Y^2(t)$). 
                 The red trajectory correspond to
                 the classical Schwarzschild solution and the 
                 green trajectory correspond to the semiclassical solution.
  }
  \end{figure}
Fig.\ref{a2} represents a plot of $X^2(t)$, in this case too the 
semiclassical solution reduces to the classical solution when $t$ approach 
the horizon but it is substantially different in the Planck region
(we recall that in the plot we have chosen $\gamma \delta \sim 1$ to amplify the
quantum correction to Schwarzschild solution 
but a semiclassical analysis is correct for $\delta \sim 10^{-33}$)\footnote{In \cite{ABM}
the spectrum of the operator $\hat{p}_c$
was calculated 
\begin{eqnarray}
\hat{p}_c |\mu, \tau \rangle = \gamma  \, l_P^2 \, \tau |\mu, \tau \rangle.
\label{pcSpe}
\end{eqnarray}
In this paper we have used dimensionless variables then 
the parameter $\delta$, which is related to the area eigenvalues 
by (\ref{pcSpe}), is order $\delta \sim 10^{-33}$.
The correct coefficient is $2 \sqrt{3}$ and it is calculated in the first of papers
\cite{ABM} comparing the area eigenvalues in the reduced Kantowski-Sachs 
model with the minimum area eigenvalue in full loop quantum gravity \cite{LoopOld}.  
}.
For the anisotropy as well as for the lapse function it is important to remember 
that the solution 
refers to the region
 $t > t^{\ast}$ while for $t = t^{\ast}$ the anisotropy 
goes towards zero, $X(t^{\ast}) \rightarrow 0$. We can conclude that for $t = t^{\ast}$
we have another event horizon, in fact for this particular value of the time coordinate 
the lapse function diverges and contemporary the anisotropy goes to zero.
This result is qualitatively similar to the modified Schwarzschild solution
obtained in asymptotic safe gravity \cite{MR} for particular values of the black hole mass
\cite{BR}.
However $t^{\ast}$ is very small 
in our semiclassical analysis and in this
region it is inevitable a complete quantum analysis of the problem as 
developed in \cite{ABM}.

\begin{figure}
 \begin{center}
  \includegraphics[height=6cm]{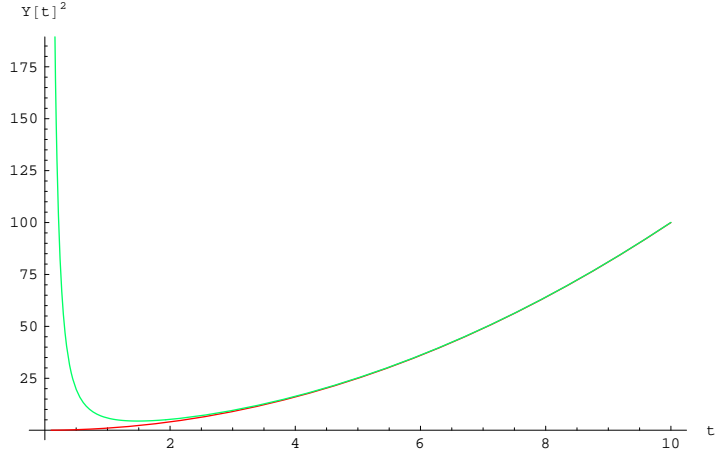}
  \end{center}
  \caption{\label{b2} Plot of $Y^2(t)$ 
 for $m = 10$ and $\gamma \delta \sim 1$
 (in the horizontal axis 
  we have the temporal coordinate $t$ and in the vertical axis we have $Y^2(t)$). 
                 The red trajectory correspond to
                 the classical Schwarzschild solution and the 
                 green trajectory correspond to the semiclassical solution.
}
  \end{figure}

The metric component $Y^2(t)$ represents the square radius of the two sphere $S^2$
and it is related to the density triad component $p_c(t)$ by the first relation reported
in (\ref{metric}). Using the solution (\ref{Sol.cpc2}) we obtain 
\begin{eqnarray}
Y^2(t) =  
\frac{1}{t^2} 
 \Big[\Big(\frac{\gamma \delta m p_b^{(0)}}{2}\Big)^2  + t^4 \Big].
 \label{Y}
\end{eqnarray}
In Fig.\ref{b2} we have a plot of $Y^2(t)$ and we can note a substantial 
difference with the classical solution. In the classical case the $S^2$ two sphere
goes to zero for $t \rightarrow 0$, in our semiclassical solution instead
the $S^2$ sphere bounces on a minimum value of the radius,
which is $Y^2(t_{min}) = \gamma \delta m$,
and it expands again to infinity for $t \rightarrow 0$. 
(we have taken the integration parameter 
$p_b^{(0)} =1$ to mach with the classical Schwarzschild solution near the horizon,
see (\ref{Sol.1}) and the first of papers \cite{ABM}).
The minimum of $Y^2(t)$ corresponds to the time coordinate 
$t_{min} = (m \gamma \delta/2)^{1/2}$
and 
$t^{\min} \gg t^{\ast}$, in fact $t^{\ast} \sim m \delta^4$ but $t_{min} \sim (m \delta)^{1/2}$, then
for $\delta \rightarrow 0$ (in the footnote one we have showed 
that $\delta \sim 10^{-33}$) we obtain $t_{\ast} \ll t_{min}$.
  
\begin{figure}
 \begin{center}
  \includegraphics[height=5cm]{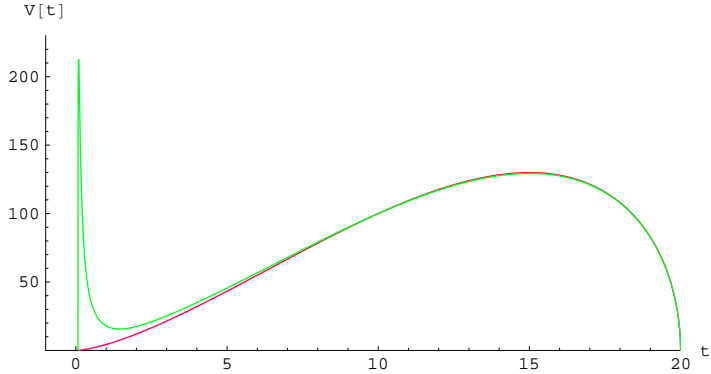}
  \end{center}
  \caption{\label{VolumeScaled} Plot of the spatial section volume $V \sim X(t) Y^2(t)$ 
 for $m = 10$ and $\gamma \delta \sim 1$
 (in the horizontal axis 
  we have the temporal coordinate $t$). 
                 The red trajectory correspond to
                 the classical volume and the 
                 green trajectory correspond to the semiclassical one.
                 From the pictures it is possible to note that the 
                 semiclassical volume (green line) is
                 zero for $t=t^{\ast}$.
                 }
  \end{figure}

In the picture Fig.\ref{VolumeScaled} we have a plot of the spatial section 
volume $V\sim X(t) Y^2(t)$ and we can see that the semiclassical 
volume has a substantially different structure at the Planck scale where it
shows a maximum for $t > t^{\ast}$ and it goes to zero
for $t = t^{\ast}$. The volume goes towards to zero on 
the event horizons but this is not a problem for the singularity 
resolution because the horizons are coordinate singularities 
and not essential singularities.

\paragraph{Quantum ambiguities and semiclassical solution.}

In this paragraph 
we want to compare the quantum spectrum of the operator $\widehat{1/|p_c|}$
with the semiclassical solution (\ref{Y}). At the quantum level the spectrum of $\widehat{1/|p_c|}$,
for a generic $SU(2)$ representation $j$ is \cite{BojInverse} 
\begin{eqnarray}
\widehat{\frac{1}{|p_{c,j}|}} |\mu, \tau \rangle = 
\left( \frac{3}{\gamma^{\frac{1}{2}} \delta \ l_P j(j+1)(2j+1)} \sum_{k = -j}^{k=j} 
\left[k \left(\sqrt{|\tau|} - \sqrt{|\tau - 2 k \delta|}\right) \right] \right)^2 |\mu, \tau \rangle.
\label{InvSpectr}
\end{eqnarray}
To compare the quantum spectrum with the semiclassical solution we must
to have a relation between the eigenvalue $\tau$ and the temporal coordinate $t$. 
We calculate this relation comparing the large $\tau$ limit of (\ref{InvSpectr}) and 
the semiclassical solution near the horizon. 
The limit of (\ref{InvSpectr}) for large eigenvalues gives
\begin{eqnarray}
\widehat{\frac{1}{|p_{c,j}|}} |\mu, \tau \rangle \rightarrow_{\tau} \frac{1}{\gamma l_P^2 |\tau|}
|\mu, \tau \rangle,
\label{InvSpectrLimit}
\end{eqnarray}
and on the other side we know that near the event horizon $1/|p_c| \rightarrow 1/t^2$, then 
comparing with (\ref{InvSpectrLimit}) we obtain $\tau=t^2/\gamma l_P^2$. At this point we
have all the ingredients to compare the quantum operator spectrum with the semiclassical
solution. 
\begin{figure}
 \begin{center}
  \includegraphics[height=8cm]{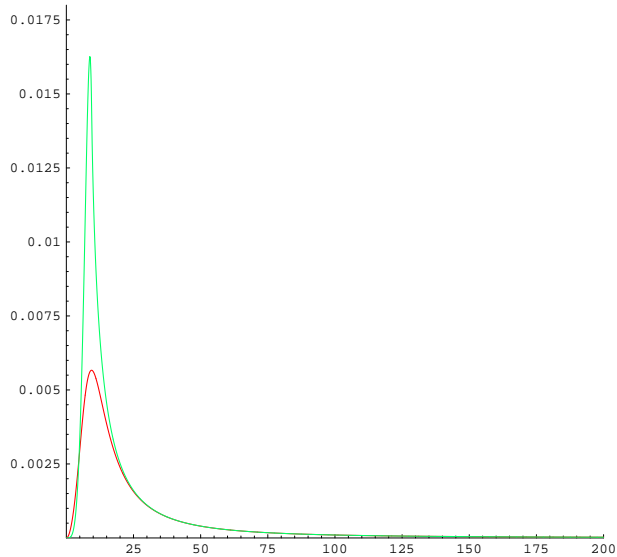}
  \end{center}
  \caption{\label{Plotjm} In this plot we compare the semiclassical solution $1/Y^2(t)$ 
  and the spectrum of the quantum operator $\widehat{1/|p_c|}$ for $j=100$ and $m =400$.
  The semiclassical solution is represented by a red line
  and the quantum spectrum by a green line.
 }
  \end{figure}
From the plot in Fig.\ref{Plotjm} it is natural to interpret the semiclassical solution
as the smooth apporoximation of the quantum operator spectrum but
the similarity between semiclassical and quantum spectrum is very
stringent only if we choose a particular relation between the black hole mass 
and the $SU(2)$ representation $j$ 
(in Fig.\ref{Plotjm} we have chosen $m=400$ and $j = 100$). 
Using an heuristic argument we can obtain the 
general relation between $m$ and $j$.
The relation is $m=4 j$ and 
now we go to show the validity of this mass
quantization formula. 

\begin{figure}
 \begin{center}
  \includegraphics[height=8cm]{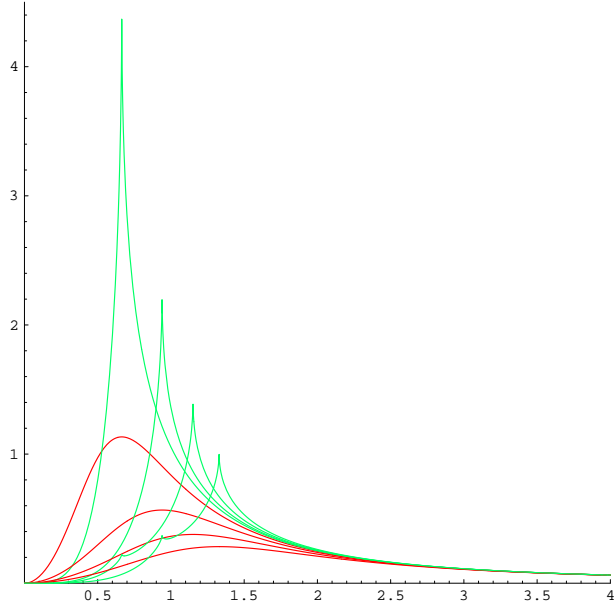}
  \end{center}
  \caption{\label{PlotYCQ} In this plot we compare the semiclassical solution $1/Y^2(t)$ 
  and the spectrum of the quantum operator $\widehat{1/|p_c|}$ for three particular 
  value of the pair $(j, m)$. From the left to the right in the plot we consider four 
  particular values of the pairs $(1/2, 2)$, $(1,4)$, $(3/2, 6)$, $(2, 8)$
  and $\gamma \delta \sim 1$. The semiclassical solution is represented by a red line
  and the quantum spectrum in green.
 }
  \end{figure}

In Fig.\ref{PlotYCQ} we have represented with a green line the quantum spectrum and 
with a red line the semiclassical solution for some values of the representation $j$
and of the mass $m$. 
This plot suggests the possibility to interpret the representation 
ambiguities  in (\ref{InvSpectr})  as a label for the mass $m$
(this idea remember a recent result about the possibility to see
ordinary matter as particular states in pure loop quantum gravity \cite{Smolin}).
In fact in the semiclassical 
solution we have a free parameter that corresponds to the black hole mass and 
on the other side in the quantum spectrum we have the representation $j$ as a free parameter.
If we interpret the semiclassical solution as the smooth approximation of the 
quantum spectrum it is possible to mach the time coordinate of the maximum 
for the two solutions. 
This is possible only if we choose a particular 
relation between $m$ and the representation $j$.  
To obtain this relation we calculate the derivative of the spectrum (\ref{InvSpectr})
respect to $\tau$ and we evaluate the derivative in 
$\tau = t^2_{min}/\gamma  =  m \delta/ 2$
($t$ is dimensionless in our analysis)
\begin{eqnarray}
\partial_{\tau} \left(\frac{1}{\sqrt{p_{\tau,j}}}\right) \Bigg|_{\tau = \frac{m \delta}{2}} = 
\frac{3}{2 \delta \ j(j+1)(2 j +1)} \sum_{k = -j}^{k = j} 
\left[k \left(\sqrt{\frac{2}{m}} - \sqrt{\frac{2}{m - 4 k}} \right) \right],
\label{DerivSpectr}
\end{eqnarray}
where $p_{\tau}$ is the eigenvalue of $\widehat{1/|p_c|}$.
Observing (\ref{DerivSpectr}) we see that in the $\widehat{1/|p_c|}$ spectrum  
the relative and absolute maximums correspond to points where the derivative 
is divergent. Those points are in $m = 4 j$ localized and this relation is also
the mass quantization formula in Planck units.
For any fixed value of the representation $j$
the classical black hole mass corresponds 
to the absolute maximum of the quantum spectrum in such representation.

\section*{Conclusions}
In this paper we have solved the Hamilton equation of motion 
for the Kantowski-Sachs space-time using the regularized Hamiltonian 
constraint suggested by loop quantum gravity.
We have obtained a solution reproducing the Schwarzschild 
solution near the event horizon but that is substantially different 
in the Planck region near the point $r=0$, where the singularity is 
(classically) localized. The structure of the solution suggests 
the possibility to have another event horizon near the point $r=0$
(this is similar to the result in ``asymptotic safety quantum gravity" \cite{BR},
but the radius of such horizon is smaller than the Planck length  
and in this region it is inevitable a complete quantum analysis 
of the problem \cite{ABM}).

Another interesting result is related to the $S^2$ sphere part
of the three metric. We obtain that 
in the semiclassical analysis the radius of the two sphere does 
not vanishes, as in the classical case, but the sphere bounces 
on a minimum radius and it expands again to infinity. 
The solution is summarized in the following table. 
\begin{center}
\begin{tabular}{|r|r|r|}
\hline
$ g_{\mu \nu} $& $\rm Semiclassical$&$ \rm Classical$\\
\hline
\hline
$- N^2(t)$ & 
$ - \frac{\gamma^2 \delta^2 \left[ \left(\frac{\gamma \delta m}{2 t^2}\right)^2 +1 \right]}{
  1 - (1 + \gamma^2 \delta^2)
\left[ \frac{\sqrt{1 + \gamma^2 \delta^2} + 1 - \left(\frac{2m}{t} \right)^{\sqrt{1 + \gamma^2 \delta^2}}
(\sqrt{1 + \gamma^2 \delta^2} - 1)}
{\sqrt{1 + \gamma^2 \delta^2} + 1 + \left(\frac{2m}{t} \right)^{\sqrt{1 + \gamma^2 \delta^2}}
(\sqrt{1 + \gamma^2 \delta^2} - 1)}
\right]^2   
}
$ & $ - \frac{1}{\frac{2m}{t} - 1}$\\
\hline
\hline
$X^2(t)$ & 
 $\frac{(2 \gamma \delta m)^2 \  
                         \left(1 - (1 + \gamma^2 \delta^2)
\left[ \frac{\sqrt{1 + \gamma^2 \delta^2} + 1 - \left(\frac{2m}{t} \right)^{\sqrt{1 + \gamma^2 \delta^2}}
(\sqrt{1 + \gamma^2 \delta^2} - 1)}
{\sqrt{1 + \gamma^2 \delta^2} + 1 + \left(\frac{2m}{t} \right)^{\sqrt{1 + \gamma^2 \delta^2}}
(\sqrt{1 + \gamma^2 \delta^2} - 1) }
\right]^2   \right) \  t^2 }{
(1 + \gamma^2 \delta^2)^2 \left(1 -
\left[ \frac{\sqrt{1 + \gamma^2 \delta^2} + 1 - \left(\frac{2m}{t} \right)^{\sqrt{1 + \gamma^2 \delta^2}}
(\sqrt{1 + \gamma^2 \delta^2} - 1)}
{\sqrt{1 + \gamma^2 \delta^2} + 1 + \left(\frac{2m}{t} \right)^{\sqrt{1 + \gamma^2 \delta^2}}
(\sqrt{1 + \gamma^2 \delta^2} - 1)}
\right]^2\right)^2
\Big[\Big(\frac{\gamma \delta m}{2}\Big)^2  + t^4 \Big]
}$
  & $\frac{2m}{t} - 1$\\
\hline
\hline
$Y^2(t)$ & $\frac{1}{t^2} \left[\left(\frac{\gamma \delta m }{2}\right)^2  + t^4\right]$ & $t^2$\\
\hline
\end{tabular}
\end{center}

Using an heuristic argument we have calculated the mass quantization
formula comparing the semiclassical and quantum spectrum of the inverse of the $S^2$ 
sphere square  
radius, $1/|p_c|$. 
Our arguments suggests the mass spectrum formula $m = 4 j$.

It is possible that the semiclassical analysis performed here will shed light on 
the problem of the ``information loss" in the process of black 
hole formation and evaporation. See in particular \cite{AB} for a possible
physical interpretation of the black hole information loss problem.



\section*{Acknowledgements}
We are grateful to Roberto Balbinot, Alfio Bonanno and  Eugenio Bianchi for many important and clarifying discussion.


\begin{thebibliography}{9}


\bibitem{book}
Carlo Rovelli, {\em Quantum Gravity}, (Cambridge University Press,
Cambridge, 2004);
A. Ashtekar, {\em Background independent quantum gravity: A Status report},
Class. Quant. Grav. 21, R53 (2004), gr-qc/0404018;
T. Thiemann, {\em Loop quantum gravity: an inside view}, hep-th/0608210; 
T. Thiemann, {\em Introduction to Modern Canonical Quantum General Relativity},
gr -qc/0110034; {\em Lectures on Loop Quantum Gravity},
Lect. Notes Phys. 631, 41-135 (2003), gr-qc/0210094

\bibitem{MR} Martin Reuter, {\em Non perturbative
evolution equation for quantum gravity}; hep-th/9605030


\bibitem{Boj} M. Bojowald, {\em Inverse scale factor in isotropic quantum geometry}, Phys. Rev. {\bf D64} 084018
(2001); M. Bojowald, {\em Loop Quantum Cosmology \textrm{IV}: discrete time evolution}, 
Class. Quant. Grav. {\bf 18}, 1071 (2001); 
Martin Bojowald,  ``Loop quantum cosmology: recent progress", gr-qc/0402053

\bibitem{MAT} A. Ashtekar, M. Bojowald and J. Lewandowski, {\em Mathematica structure of loop quantum cosmology}, Adv. Theor. Math. Phys. {\bf 7} (2003) 233-268, gr-qc/0304074

\bibitem{work1} Leonardo Modesto, {\em Disappearance of the black hole singularity in loop  quantum gravity}, Phys. Rev. D 70 (2004) 124009, gr-qc/0407097

\bibitem{work2} Leonardo Modesto, {\em The kantowski-Sachs space-time in loop quantum gravity}, International Journal of Theoretical Physics, published on line 1 june 2006, 
 gr-qc/0411032
 
\bibitem{work3} Leonardo Modesto, {\em Gravitational collapse in loop quantum gravity}, 
gr-qc/0610074;
Leonardo Modesto, {\em Quantum gravitational collapse}, gr-qc/0504043

\bibitem{ABM}  A. Ashtekar and M. Bojowald, 
{\em Quantum geometry and Schwarzschild singularity} Class. Quant. Grav. 23 (2006) 391-411,
gr-qc/0509075;  Leonardo Modesto, {Loop quantum black hole}, Class. Quant. Grav. 23 (2006) 5587-5602, gr-qc/0509078

\bibitem{BR} Alfio Bonanno, Martin Reuter, {\em Renormalization group improved black hole space-times}, Phys. Rev. D 62 (2000) 043008, hep-th/0002196;
Alfio Bonanno, Martin Reuter 
{\em Spacetime structure of an evaporating black hole in quantum gravity},  
Phys. Rev. D 73 (2006) 083005,
hep-th/0602159

\bibitem{NC} Christian G. Boehmer and Kevin Vandersloot, 
{\em Loop Quantum Dynamics of the Schwarzschild Interior},
Phys. Rev. D76 (2007) 104030,  
arXiv:0709.2129

\bibitem{KS} R. Kantowski and R. K. Sachs, J. Math. Phys. {\bf 7} (3) (1966)


\bibitem{variables} Abhay Ashtekar, {\em New Hamiltonian formulation of general relativity}, 
Phys. Rev. D {\bf 36} 1587-1602


\bibitem{BojThiemann} I. Bengtsson,  ``Note on Ashtekar's variables in the spherically
symmetric case", Class. Quant. Grav. {\bf 5} (1988) L139-L142;
I. Bengtsson,  ``A new phase for general relativity?", Class. Quant. Grav. {\bf 7} (1990) 27-39;
H.A. Kastrup \& T. Thiemann,  ``Spherically symmetric gravity as a complete integrable system",
Nucl. Phys. B 425 (1994) 665-686, gr-qc/9401032;
M. Bojowald \& H.A. Kastrup,   ``Quantum symmetry reduction of Diffeomorphism invariant
theories of connections", JHEP 0002 (2000) 030, hep-th/9907041;
L. Bombelli \& R. J. Torrence,  ``Perfect fluids and Ashtekar variables,
with application to Kantowski-Sachs models", Class. Quant. Grav. {\bf 7} (1990) 1747-1745; 
M. Bojowald,  ``Spherically symmetric quantum geometry: states and basic operators",
Class. Quant. Grav. {\bf 21} (2004) 3733-3753, gr-qc/0407017; 
M. Bojowald and R. Swiderski,  ``Spherically symmetric quantum geometry: Hamiltonian Constraint",
Class. Quant. Grav. {\bf 23} (2006) 2129-2154, gr-qc/0511108


 
 \bibitem{LoopOld}  C.~Rovelli and L.~Smolin,
  ``Loop Space Representation Of Quantum General Relativity,''
  Nucl.\ Phys.\ B {\bf 331} (1990) 80; 
   C.~Rovelli and L.~Smolin,
  ``Discreteness of area and volume in quantum gravity,''
  Nucl.\ Phys.\ B {\bf 442} (1995) 593

 
 










\bibitem{BojInverse} Martin Bojowald, {\em Quantization ambiguities in isotropic 
quantum geometry}; gr-qc/0206053

\bibitem{AB} Abhay Ashtekar \& Martin Bojowald, {\em Black hole evaporation : A paradigm}
Class. Quant. Grav. {\bf 22} (2005) 3349-3362, gr-qc/0504029

\bibitem{Smolin}  Sundance O. Bilson-Thompson,  Fotini Markopoulou and Lee Smolin,
{\em Quantum gravity and the standard model}, hep-th/0603022

\end{thebibliography}
\end{document}